\pdfoutput=1
\RequirePackage{ifpdf}
\ifpdf 
\documentclass[pdftex]{sigma}
\else
\documentclass{sigma}
\fi

\begin{document}

\allowdisplaybreaks

\renewcommand{\PaperNumber}{008}

\FirstPageHeading

\ShortArticleName{Discrete Spectral Transformations of SOPs and Associated DIS}

\ArticleName{Discrete Spectral Transformations \\ of Skew Orthogonal Polynomials\\ and Associated Discrete Integrable Systems}

\Author{Hiroshi MIKI, Hiroaki GODA and Satoshi TSUJIMOTO}
\AuthorNameForHeading{H.~Miki, H.~Goda and S.~Tsujimoto}

\Address{Department of Applied Mathematics and Physics, Graduate School of Informatics,\\ Kyoto University, Sakyo-Ku, Kyoto 606 8501, Japan}

\Email{\href{mailto:miki@amp.i.kyoto-u.ac.jp}{miki@amp.i.kyoto-u.ac.jp}, \href{mailto:tujimoto@i.kyoto-u.ac.jp}{tujimoto@i.kyoto-u.ac.jp}}
\URLaddress{\url{http://www-is.amp.i.kyoto-u.ac.jp/lab/en/miki/}}

\ArticleDates{Received December 01, 2011, in f\/inal form February 20, 2012; Published online February 29, 2012}

\Abstract{Discrete spectral transformations of skew orthogonal polynomials are presented. From these spectral transformations, it is shown that
the corresponding discrete integrable systems are derived both in $1+1$ dimension and in $2+1$ dimension. Especially in the $(2+1)$-dimensional case,  the corresponding system can be extended to
$2\times 2$ matrix form. The factorization theorem of the Christof\/fel kernel for skew orthogonal polynomials in random matrix theory is presented as a by-product of these transformations.}


\Classification{42C05; 35C05; 37K60; 15B52}

\section{Introduction}

Orthogonal polynomials and their varieties have been investigated for a long time and yielded a lot of contributions to various areas including quantum physics, random matrices, wavelets and so on.
Over the last twenty years, several integrable systems, both in continuous variable and in discrete variable, have been found to relate to such polynomials through their spectral transformations. One of the well-known results is the relation between ordinary orthogonal polynomials (OPs) and Toda chain (see~\cite{Aptekarev,Spiridonov2} for further details). For recent years, from this point of view, new types of integrable systems have been derived~\cite{Spiridonov1,Spiridonov3} and the algebraic structures of their solutions also have been discussed in \cite{Mukaihira,Mukaihira2}.

In this paper, we treat the skew orthogonal polynomials (SOPs), f\/irst introduced in the theo\-ry of random matrices \cite{Dyson}.
It is known that the continuous spectral transformation of SOPs induces the semi-discrete integrable system, known as Pfaf\/f lattice \cite{Adler,Adler3}, and its relations to the Lie algebra, geometry and matrix eigenvalue problems are discussed in~\cite{Adler,Kodama2,Kodama} respectively. Although, the full-discrete integrable system associated with SOPs, namely the discrete counterpart of Pfaf\/f lattice, still remains to be found and even the discrete spectral transformation of SOPs is not obtained.
Recalling the application of the discrete integrable system to several areas such as eigenvalue problems and so on (e.g.~\cite{Minesaki,Tsujimoto}), the discrete integrable system associated with SOPs, if obtained, can be expected to have an application or contribution to such areas.  We shall derive the corresponding full-discrete integrable systems from their discrete spectral transformations.

This paper is organized as follows. In Section~\ref{section2}, a brief review of the relationship between OPs and the discrete time Toda equation is given and we propose the discrete spectral transformation of SOPs. From these transformations, we derive the corresponding discrete integrable system, which takes $(1+1)$-dimensional form. In Section~\ref{section3}, another discrete integrable system associated with SOPs is shown to be derived also from their discrete spectral transformations, which takes $(2+1)$-dimensional form. In Section~\ref{section4}, we also show that the system obtained in Section~\ref{section3} can be extended to $2 \times 2$ matrix form. In Section~\ref{section5}, concluding remarks are presented.

\section{Skew orthogonal polynomials}\label{section2}

We f\/irst brief\/ly review a relation between the discrete Toda time equation and the discrete spectral transformation of OPs according to \cite{Spiridonov2}.
The discrete time Toda equation (on semi-inf\/inite lattice)
\begin{gather*}
 U_n^tV_n^t=U_{n-1}^{t+1}V_{n}^{t+1},\qquad
 U_n^t+V_{n+1}^t=U_n^{t+1}+V_n^{t+1},\qquad V_0^t=0,
 \end{gather*}
is known to admit the following discrete Lax pair:
\begin{gather}\label{dlaxToda}
x\Phi^{t+1}=
\begin{pmatrix}
U_0^t & 1   &     &   & \\
  & U_1^t & 1   &   & \\
    &  & U_2^t & 1 & \\
    &     &  & \ddots & \ddots
\end{pmatrix}\Phi^t, \qquad \Phi^{t}=
\begin{pmatrix}
1 &    &     &   & \\
V_1^t  & 1 &    &   & \\
    & V_2^t  & 1 &  & \\
    &     & \ddots   & \ddots &
\end{pmatrix}\Phi^{t+1},
\end{gather}
where $x$ is a spectral parameter and $\Phi^t =(\phi_0^t(x),\phi_1^t(x),\dots )^T$ is a wave function. From \eqref{dlaxToda}, we obtain
\begin{gather*}
x\Phi^{t+1}=
\begin{pmatrix}
U_0^{t}+V_1^t  & 1   &     &   & \\
U_1^tV_1^t & U_1^{t}+V_2^t  & 1   &   & \\
    & U_2^tV_2^t & U_2^t+V_3^t  & 1 & \\
    &     & \ddots & \ddots & \ddots
\end{pmatrix}\Phi^{t+1},
\end{gather*}
for all $t$. If we normalize $\phi_0^t(x)=1$, Favard's theorem shows that $\{ \phi_n^t(x)\}_{n=0}^{\infty }$ is a sequence of monic orthogonal polynomials, i.e. there exists some linear functional $\mathcal{L}^t:\mathbb{R}[x]\rightarrow \mathbb{R}$ such that
\begin{gather*}
\mathcal{L}^t[\phi _m^t(x)\phi_n^t(x)]=h_n^{t}\delta_{mn},\qquad h_n^t\ne0.
\end{gather*}
Therefore, one can easily f\/ind that the discrete Lax pair \eqref{dlaxToda} is just the discrete spectral transformation of orthogonal polynomials known as Christof\/fel transformation and Geronimus transformation \cite{Chihara,Geronimus}, respectively.

Our main aim is to derive the discrete integrable systems associated with SOPs.
To that end, we shall give the discrete spectral transformations of SOPs like Christof\/fel and Geronimus transformation. We f\/irst begin with the def\/inition of SOPs:

\begin{definition}
Let $\langle \cdot |\cdot \rangle $ be the bilinear 2-form from $\mathbb{R}[z]\times \mathbb{R}[z]$ to $\rightarrow \mathbb{R}$ which satisfy the skew symmetric relation
\begin{gather*}
\langle f(z) | g(z) \rangle = -\langle g(z) | f(z) \rangle.
\end{gather*}
This skew-symmetric bilinear 2-form is often called ``skew-inner product''.
The SOPs $\{ q_{n}(z)\}_{n=0}^{\infty }$ with respect to the skew-inner product $\langle \cdot | \cdot \rangle$ are def\/ined as the set of polynomials satisfying the following relations:
\begin{gather}
 \langle q_{2m}(z) |q_{2n+1}(z) \rangle = r_n\delta_{mn},\qquad r_n\ne 0, \nonumber\\
 \langle q_{2m}(z) | q_{2n}(z) \rangle=-\langle q_{2m+1}(z) | q_{2n+1}(z) \rangle=0,\label{eq1}
\end{gather}
for all nonnegative integers $m$, $n$.
\end{definition}
From now on, unless specif\/ied, we assume that SOPs are monic, which ensures the uniqueness of SOPs of even degree.
However, SOPs of odd degree have the following ambiguity: the skew orthogonality relation \eqref{eq1} is invariant under the replacement
\begin{gather}\label{ambiguity}
q_{2n+1}(z)\rightarrow q_{2n+1}(z)+\alpha_nq_{2n}(z) ,
\end{gather}
for any $\alpha _n\in \mathbb{R}$.  In many cases, the coef\/f\/icient of $z^{2n}$ in $q_{2n+1}(z)$ is usually chosen as $0$ and then the SOPs are uniquely determined.

The discrete spectral transformation of SOPs (that is, the transformation from SOPs to another SOPs) can be constructed.
\begin{theorem}\label{theorem1}
Let $\{ q_n(z)\}_{n=0}^{\infty }$ be SOPs with respect to the skew-inner product $\langle \cdot | \cdot \rangle$. Let us denote some constant parameter by $\lambda $ satisfying
\begin{gather*}
\lambda \in \{ z \in \mathbb{C} ~|~q_{2n}(z)\ne 0,\quad \textrm{for all } n=0,1,\dots \}.
\end{gather*}
Consider the new sequence of polynomials $\{ q_{n}^*(z)\}_{n=0}^{\infty }$ defined by
\begin{gather}
q_{2n}^*(z) =\frac{1}{z-\lambda }\left( \sum_{k=0}^n \frac{r_n}{r_k} \cdot \frac{q_{2k+1}(z)q_{2k}(\lambda ) -q_{2k+1}(\lambda ) q_{2k}(z)}{q_{2n}(\lambda )}\right) , \label{eq2}\\
q_{2n+1}^*(z) =\frac{1}{z-\lambda }\left( q_{2n+2}(z)-\frac{q_{2n+2}(\lambda )}{q_{2n}(\lambda )} q_{2n}(z)\right) + \alpha_n q_{2n}^*(z) \label{eq4},
\end{gather}
where $\alpha_n$ is an arbitrary parameter. Then $\{ q_{n}^*(z)\}_{n=0}^{\infty }$ are again SOPs with respect to the modified skew-inner product
\begin{gather}\label{msip}
\langle \cdot | \cdot \rangle^*:=\langle (z-\lambda ) \cdot | (z-\lambda )\cdot \rangle .
\end{gather}
\end{theorem}
\begin{proof}
In this proof, because of the equation \eqref{ambiguity}, we take the arbitrary parameter $\alpha_n=0$ for all~$n$ in~\eqref{eq4} for simplicity. In order to prove the theorem, it is enough to check the skew orthogonal relation \eqref{eq1} for  $\{ q_{n}^*(z)\}_{n=0}^{\infty }$.
Using the skew orthogonality relation \eqref{eq1} for  $\{ q_{n}(z)\}_{n=0}^{\infty }$, we f\/ind
\begin{gather*}
 \langle q_{2n}^*(z),q_{2m+1}^*(z)\rangle^*
 =\langle (z-\lambda )q_{2n}^*(z),(z-\lambda )q_{2m+1}^*(z)\rangle\nonumber\\
 \qquad{} =\left\langle \sum_{k=0}^n\frac{q_{2k}(\lambda )q_{2k+1}(z)-q_{2k+1}(\lambda )q_{2k}(z)}{r_kr_n^{-1}q_{2n}(\lambda )}\right| \left. q_{2m+2}(z)-\frac{q_{2m+2}(\lambda )}{q_{2m}(\lambda )}q_{2m}(z)\right\rangle\nonumber\\
\qquad{} =\frac{r_n}{q_{2n}(\lambda )}\sum_{k=0}^n\left\langle \frac{q_{2k}(\lambda )q_{2k+1}(z)}{r_k}\right| \left. q_{2m+2}(z)-\frac{q_{2m+2}(\lambda )}{q_{2m}(\lambda )}q_{2m}(z)\right\rangle
 =\frac{q_{2n+2}(\lambda )}{q_{2n}(\lambda )}r_n\delta_{mn}.
\end{gather*}
Similarly, direct calculations also show us
\begin{gather*}
\langle q_{2m}^*(z),q_{2n}^*(z)\rangle^*=\langle q_{2n+1}^*(z),q_{2m+1}^*(z)\rangle^*=0.
\end{gather*}
This completes the proof.
\end{proof}

Now observing the discrete spectral transformation of even-degree SOPs \eqref{eq2}, one can easily f\/ind that except for the multiplier factor, this is equivalent to the Christof\/fel--Darboux kernel for skew orthogonal polynomials (we will call this ``skew-Christof\/fel kernel'') \cite{Ghosh,Mehta} introduced in the theory of random matrices which takes the form
\begin{gather}\label{eq3}
I_{N}(x,y)=\sum_{k=0}^N \frac{q_{2k+1}(x)q_{2k}(y ) -q_{2k+1}(y ) q_{2k}(x)}{r_k},
\end{gather}
where  $\{ q_{n}(z)\}_{n=0}^{\infty }$ are some SOPs.
We hence obtain the following corollary:
\begin{corollary}
The skew-Christoffel kernel \eqref{eq3} can be factorized into the two SOPs:
\begin{gather}\label{sckernel}
I_{N}(x,y)=(x-y)\frac{q_{2N}(x)q_{2N}^*(x)}{r_N},
\end{gather}
where the SOPs $q_{2N}(x)$ and $q_{2N}^*(x)$ are connected to each other by the spectral transforma\-tion~\eqref{eq1} with $z=x$, $\lambda =y$.
\end{corollary}

There exist many works on the skew-Christof\/fel kernel for the specif\/ic weight cases (see e.g.~\cite{Adler2,Ghosh}).
Here we stress that the factorization \eqref{sckernel} holds not only for the case mentioned above but also for any well-def\/ined skew inner product.  Especially, consider the skew inner product in orthogonal and symplectic ensemble given by:
\begin{gather}
 \langle f(z)|g(z)\rangle_{\textrm{orthogonal}}=\iint_{\mathbb{D}^2} \mathrm{sgn} (z_1-z_2)f(z_1)g(z_2)w(z_1)w(z_2)dz_1 dz_2, \nonumber\\
 \langle f(z)|g(z)\rangle_{\textrm{symplectic}}=\int_{\mathbb{D}} (f(z)g'(z)-f'(z)g(z))w(z)dz,
\end{gather}
where $f'(z)=\frac{d}{dz}f(z)$ and $\mathbb{D}$, $w(z)$ are the corresponding integral domain and weight function respectively. For such skew inner product, the modif\/ied skew inner product \eqref{msip} can be written as follows:
\begin{gather*}
 \langle (z-\lambda )f(z)|(z-\lambda )g(z)\rangle_{\textrm{orthogonal}}=\iint_{\mathbb{D}^2} \mathrm{sgn}(z_1-z_2) f(z_1)g(z_2)\bar{w}(z_1)\bar{w}(z_2)dz_1dz_2, \nonumber\\
 \langle (z-\lambda )f(z)|(z-\lambda ) g(z)\rangle_{\textrm{symplectic}}=\int_{\mathbb{D}} (f(z)g'(z)-f'(z)g(z))\bar{w}(z)dz,
\end{gather*}
with $\bar{w}(z)=(z-\lambda )w(z)$. This shows that the skew-Christof\/fel kernel in random matrix theory can be factorized into two even-degree SOPs with respect to the dif\/ferent weight function.

\begin{remark}
Similar relations hold between the Christof\/fel transformation of OPs and the Christof\/fel kernels.
Hence we call the discrete spectral transformation \eqref{eq2}, \eqref{eq4} as the skew-Christof\/fel transformation.
\end{remark}

Iterating the skew-Christof\/fel transformation, we obtain a sequence of SOPs by the following procedure:
\begin{gather*}
q_n^{0}(z) :=q_{n}(z),\nonumber\\
q_{2n}^{t+1}(z) :=\frac{1}{z-\lambda }\left( q_{2n+1}^t(z)+\sum_{k=0}^nA_{n,k}^tq_{2k}^t(z)+\sum_{k=0}^{n-1}B_{n,k}^tq_{2k+1}^t(z)\right), \nonumber\\
q_{2n+1}^{t+1}(z) :=\frac{1}{z-\lambda }\left( q_{2n+2}^t(z)+C_n^tq_{2n}^t(z)\right) + \alpha_n^t q_{2n}^{t+1}(z),
\end{gather*}
where
\begin{gather*}
A_{n,k}^t =\frac{r_n^t}{r_k^t}\cdot \frac{q_{2k}^t(\lambda )}{q_{2n}^t(\lambda )},\qquad B_{n,k}^t=-\frac{r_n^t}{r_k^t}\cdot \frac{q_{2k+1}^t(\lambda )}{q_{2n}^t(\lambda )},\qquad
C_n^t=-\frac{q_{2n+2}^t(\lambda )}{q_{2n}^t(\lambda )}.
\end{gather*}
Here we denote the corresponding skew-inner product by $\langle \cdot |\cdot \rangle^t$. Then, skew-orthogonality relations are given as follows:{\samepage
\begin{gather}
 \langle \cdot |\cdot \rangle^0:=\langle \cdot | \cdot \rangle,\qquad
 \langle \cdot |\cdot \rangle^{t+1}:=\langle (z-\lambda )\cdot | (z-\lambda ) \cdot \rangle^t,\nonumber\\
 \langle q_{2m}^t(z),q_{2n+1}^t(z)\rangle^t = r_n^t\delta_{mn},\qquad
 \langle q_{2m}^t(z),q_{2n}^t(z)\rangle ^t=\langle q_{2m+1}^t(z),q_{2n+1}^t(z)\rangle ^t=0.\label{dis}
\end{gather}
For simplicity, we assume $\alpha_n^t=0$.}

We then proceed to f\/ind the inverse transformation of the skew-Christof\/fel transformation. Using the skew-orthogonality relation \eqref{eq1}, we can indeed construct the dif\/ferent contiguous relations between $\{ q_{n}^t(z)\} _{n=0}^{\infty }$ and $\{ q_{n}^{t+1}(z)\}_{n=0}^{\infty }$ as follows:
\begin{gather}
q_{2n}^t(z) =q_{2n}^{t+1}(z)+\sum_{k=0}^{n-1}\alpha_{n,k}^tq_{2k}^{t+1}(z)+\sum_{k=0}^{n-1}\beta_{n,k}^tq_{2k+1}^{t+1}(z),\nonumber\\
q_{2n+1}^t(z) =q_{2n+1}^{t+1}(z)+\sum_{k=0}^{n}\gamma_{n,k}^tq_{2k}^{t+1}(z)+\sum_{k=0}^{n-1}\epsilon_{n,k}^tq_{2k+1}^{t+1}(z),\label{gero}
\end{gather}
where
\begin{gather*}
 \alpha_{n,k}^t=\frac{1}{r_k^{t+1}}\langle (z-\lambda )q_{2n}^{t+1}(z)|(z-\lambda ) q_{2k+1}^{t+1}(z)\rangle^t,\nonumber\\
 \beta_{n,k}^t=\frac{1}{r_k^{t+1}}\langle (z-\lambda )q_{2k}^{t+1}(z)|(z-\lambda )q_{2n}^{t+1}(z)\rangle^t ,\nonumber\\
 \gamma_{n,k}^t=\frac{1}{r_k^{t+1}}\langle (z-\lambda )q_{2n+1}^t(z)| (z-\lambda )q_{2k+1}^{t+1}(z)\rangle^t,\nonumber\\
  \epsilon_{n,k}^t=\frac{1}{r_k^{t+1}}\langle (z-\lambda )q_{2k}^{t+1}(z)|(z-\lambda )q_{2n+1}^t(z)\rangle^t.
\end{gather*}
Combining \eqref{dis} and \eqref{gero}, we can form the following Lax pair:
\begin{gather}
(z-\lambda) \Phi ^{t+1} =L^t\Phi^t
:=
\begin{pmatrix}
A_{0,0}^t  & 1  &  &  & &\\
C_0^t  & 0  & 1 &   & &\\
 A_{1,0}^t   &  B_{1,0}^t  & A_{1,1}^t& 1  & &\\
 0  & 0  &  C_1^t  & 0 & 1   &  \\
 \vdots & \ddots & \ddots & \ddots & \ddots & \ddots
\end{pmatrix}\Phi^t, \nonumber\\
\Phi^t =R^t\Phi^{t+1}
:=\begin{pmatrix}
1 & &  & & \\
\gamma_{0,0}^t & 1 &  &  & \\
\alpha_{1,0}^t & \beta_{1,0}^t & 1 & & \\
\gamma_{1,0}^t & \epsilon_{1,0}^t & \gamma_{1,1}^t & 1 & \\
\vdots & \ddots & \ddots & \ddots & \ddots
\end{pmatrix}
\Phi^{t+1},\label{1+1}
\end{gather}
where the wave-function $\Phi^t =(q_{0}^t(z),q_1^t(z),q_2^t(z),\dots )^T$ is the sequence of SOPs.
From the compatibility condition of \eqref{1+1}, the discrete Lax equation is derived:
\begin{gather}\label{d-lax}
L^tR^t=R^{t+1}L^{t+1}.
\end{gather}
We can observe that if we take the suitable limit, the second equation of \eqref{dis} goes to the relation:
\begin{gather*}
\frac{d}{dt}\langle \cdot | \cdot \rangle=\langle z \cdot | \cdot \rangle + \langle  \cdot | z \cdot \rangle,
\end{gather*}
which appears in the Pfaf\/f lattice proposed by Adler et al.~\cite{Adler}.
The discrete Lax equation \eqref{d-lax} thus can be considered to be  the discrete counterpart of the Pfaf\/f lattice.

\section[Discrete integrable systems of SOPs in $2+1$ dimension]{Discrete integrable systems of SOPs in $\boldsymbol{2+1}$ dimension}\label{section3}

In the previous section, we derive the discrete integrable system \eqref{d-lax} from the discrete spectral transformation of SOPs.
However the system is \emph{non-local} and therefore complicated. The similar problem occurs in the case of ordinary biorthogonal polynomials and discrete two-dimensional Toda (discrete KP) equation. This problem can be solved by introducing ``another time-evolution'' \cite{Tsujimoto2}. Furthermore, observing the relation between skew-Christof\/fel kernel and SOPs \eqref{sckernel}, one can realize that only even-degree SOPs appear. We hence expect that even-degree SOPs play an crucial role in the corresponding discrete integrable systems. Due to the guiding principles mentioned above, we shall construct the discrete integrable systems in $(2+1)$-dimensional form associated with only even-degree SOPs.

Let $\langle \cdot |\cdot \rangle^{s,t}$ be a skew-inner product such that
\begin{gather}
\langle \cdot |\cdot \rangle^{s+1,t} =\langle (z-\mu )\cdot |(z-\mu )\cdot \rangle^{s,t},\qquad
\langle \cdot |\cdot \rangle^{s,t+1} =\langle (z-\lambda )\cdot |(z-\lambda )\cdot \rangle^{s,t},\label{2te}
\end{gather}
where we assume $\lambda \ne \mu$. We denote the corresponding SOPs by $\{ q_{n}^{s,t}(z)\}_{n=0}^{\infty }$. From Theorem \ref{theorem1}, $\{ q_{n}^{s,t}(z)\}_{n=0}^{\infty }$ and  $\{ q_{n}^{s+1,t}(z)\}_{n=0}^{\infty }$ are connected by the relation~\eqref{eq2} and \eqref{eq4} with $\lambda =\mu$. Similarly, $\{ q_{n}^{s,t}(z)\}_{n=0}^{\infty }$ are also related to $\{ q_{n}^{s,t+1}(z)\}_{n=0}^{\infty }$ by~\eqref{eq2} and~\eqref{eq4}.

We shall f\/ind the local contiguous relations among $\{ q_{n}^{s,t}(z)\}_{n=0}^{\infty }$. To that end, we f\/irst recall the Pfaf\/f\/ian expression of ordinary SOPs \cite{Adler}, which will play a crucial role in the following sections. Pfaf\/f\/ians are def\/ined by:
\begin{gather*}
 \mathrm{Pf}(i_0,\dots ,i_{2n-1}):=\sum_{\sigma \in \mathfrak{S}_{2n}}\frac{\textrm{sgn}\, \sigma }{n!2^n}\prod _{0\le i\le n-1}\mathrm{Pf}(i_{\sigma (2i)},i_{\sigma (2i+1)}),
\end{gather*}
where the elements of Pfaf\/f\/ians $\mathrm{Pf}(i,j)$ are supposed to satisfy the skew-symmetric relation:
\begin{gather}\label{antisym}
\mathrm{Pf}(i,j)=-\mathrm{Pf}(j,i).
\end{gather} From the def\/inition of Pfaf\/f\/ians, we can f\/ind the following properties of Pfaf\/f\/ians which resemble those of determinants:
\begin{gather}
 \mathrm{Pf}(i_0,\dots ,i_{2n-1})=\sum_{k=0}^{2n-2}(-1)^k(i_k,i_{2n-1})(i_0,\dots, \widehat{i_k},\dots ,i_{2n-2}), \label{anotherPfaff} \\
\mathrm{Pf}(i_0,\dots ,a\cdot i_{a}+b\cdot i_{b},\dots ,i_{2n-1})\nonumber\\
\qquad{}
= a\mathrm{Pf}(i_0,\dots ,i_a ,\dots ,i_{2n-1})
   +b\mathrm{Pf}(i_0,\dots ,i_b,\dots ,i_{2n-1}), \label{linear:pfaff}
\\
\mathrm{Pf}(\dots, s,\dots ,s,\dots )=0, \label{zero:pfaff}
\end{gather}
where $\widehat{j}$ means the deletion of $j$ and
\begin{gather}\label{lpf}
\mathrm{Pf}(i,a\cdot i_{a}+b\cdot i_{b}):=a\mathrm{Pf}(i,i_a)+b\mathrm{Pf}(i,i_b).
\end{gather}
 For further details of Pfaf\/f\/ians, see e.g.~\cite{Knuth,Ohta}.

Employing the ``skew'' moments $\langle z^i| z^j\rangle$ as the elements of Pfaf\/f\/ians, SOPs can be written in terms of Pfaf\/f\/ians as well as ordinary OPs in terms of determinants.
\begin{proposition}[\protect{\cite{Adler}}]\label{Pfaffian expression of SOPs}
The SOPs with respect to $\langle \cdot |\cdot \rangle$ can be expressed in terms of Pfaffians:
\begin{gather*}
q_{2n}(z) =\frac{\mathrm{Pf}(0,1,\dots ,2n,z)}{\mathrm{Pf}(0,1,\dots ,2n-1)},\\
q_{2n+1}(z) =q_{2n}(z)=\frac{\mathrm{Pf}(0,1,\dots ,2n-1,2n+1,z)}{\mathrm{Pf}(0,1,\dots ,2n-1)}+\alpha_n q_{2n}(z),
\end{gather*}
where $\alpha_n$ is an arbitrary parameter and the elements of Pfaffians are defined by
\begin{gather*}
\mathrm{Pf}(i,j)=\langle z^i|z^j\rangle,\qquad \mathrm{Pf}(i,z)=z^i
\end{gather*}
for all nonnegative integers $i$, $j$.
\end{proposition}

Extending this proposition, we have obtained the Pfaf\/f\/ian expressions of the time-evolved even degree SOPs.
\begin{theorem}\label{pfexpress}
The even degree SOPs $\{ q_{2n}^{s,t}(z)\}_{n=0}^{\infty }$,$\{ q_{2n}^{s+1,t}(z)\}_{n=0}^{\infty }$,$\{ q_{2n}^{s,t+1}(z)\}_{n=0}^{\infty }$,
$\{ q_{2n}^{s+1,t+1}(z)\}_{n=0}^{\infty }$ can be written in terms of the moments with respect to $\langle \cdot | \cdot \rangle^{s,t}$:
\begin{gather}
q_{2n}^{s,t}(z) =\frac{\mathrm{Pf}(0,1,\dots ,2n,z)}{\mathrm{Pf}(0,1,\dots ,2n-1)},\nonumber\\
q_{2n}^{s+1,t}(z) =\frac{1}{z-\mu }\cdot \frac{\mathrm{Pf}(0,1,\dots ,2n+1,\mu ,z)}{\mathrm{Pf}(0,1,\dots ,2n,\mu )},\nonumber\\
q_{2n}^{s,t+1}(z) =\frac{1}{z-\lambda }\cdot \frac{\mathrm{Pf}(0,1,\dots ,2n+1,\lambda ,z)}{\mathrm{Pf}(0,1,\dots ,2n,\lambda )},\nonumber\\
q_{2n}^{s+1,t+1}(z) =\frac{1}{(z-\mu )(z-\lambda )}\cdot \frac{\mathrm{Pf}(0,1,\dots ,2n+2,\mu ,\lambda ,z)}{\mathrm{Pf}(0,1,\dots ,2n+1,\mu, \lambda )},\label{tesops}
\end{gather}
where the elements of Pfaffians are defined by
\begin{gather}
 \mathrm{Pf}(i,j)=s_{i,j}^{s,t}:=\langle z^i|z^j\rangle^{s,t},\nonumber\\
 \mathrm{Pf}(i,z)=z^i,\qquad \mathrm{Pf}(i,\lambda )=\lambda ^i,\qquad \mathrm{Pf}(i,\mu) =\mu ^i,\nonumber\\
 \mathrm{Pf}(z,\mu )=\mathrm{Pf}(\mu,\lambda )=\mathrm{Pf}(\lambda ,z)=0.\label{pfelements}
\end{gather}
\end{theorem}
Before proving this theorem, we show the properties of the Pfaf\/f\/ian elements.
\begin{lemma}\label{lemma1}
Employing the notation \eqref{lpf} in \eqref{linear:pfaff}, for the elements of the Pfaffians defined by~\eqref{pfelements}, the following properties hold:
\begin{gather*}
 \mathrm{Pf}((i+1)-\mu \cdot i, (j+1)-\mu \cdot j)  =\mathrm{Pf}(i+1,j+1)-\mu \mathrm{Pf}(i+1,j)-\mu \mathrm{Pf}(i,j)+\mu ^2\mathrm{Pf}(i,j),\! \nonumber\\
 \mathrm{Pf}((i+1)-\mu\cdot i,\mu )=0.
\end{gather*}
\end{lemma}
\begin{proof}
Following the def\/inition \eqref{lpf} and anti-symmetric property of Pfaf\/f\/ian elements \eqref{antisym}, one can f\/ind
\begin{gather*}
 \mathrm{Pf}((i+1)-\mu \cdot i, (j+1)-\mu \cdot j)
 =\mathrm{Pf}((i+1)-\mu \cdot i, j+1)-\mu \mathrm{Pf}((i+1)-\mu \cdot i,j)\nonumber \\
 \qquad{} =-\mathrm{Pf}( j+1,(i+1)-\mu \cdot i,)+\mu \mathrm{Pf}(j,\mu (i+1)-\mu \cdot i) \nonumber\\
 \qquad{} =-\mathrm{Pf}(j+1,i+1)-\mu \mathrm{Pf}(j+1,i)+\mu \mathrm{Pf}(j,i+1)-\mu ^2 \mathrm{Pf}(j,i) \nonumber\\
 \qquad{} =\mathrm{Pf}(i+1,j+1)-\mu \mathrm{Pf}(i+1,j)-\mu \mathrm{Pf}(i,j)+\mu ^2\mathrm{Pf}(i,j).
\end{gather*}
One can also f\/ind
\begin{gather*}
 \mathrm{Pf}(i+1-\mu \cdot i,\mu )
 =\mathrm{Pf}(i+1,\mu) -\mu \mathrm{Pf}(i,\mu )=\mu^{i+1}-\mu^{i+1}=0.
\end{gather*}
This completes the proof.
\end{proof}

\begin{proof}[Proof of Theorem~\ref{pfexpress}]
We introduce the function $\tau _{n}^{s,t}$ and $\hat{\tau}_{n}^{s,t}$ as follows:
\begin{gather}
\tau _n^{s,t} =\mathrm{Pf}(0,\dots ,2n-1),\qquad
\hat{\tau} _n^{s,t} =\mathrm{Pf}(0,\dots ,2n,z).\label{tau}
\end{gather}
It is also convenient to introduce the shift operator $T_s$ and $T_t$ satisfying
\begin{gather*}
T_s[f(s)]=f(s+1),\qquad T_t[f(t)]=f(t+1)
\end{gather*}
and the notation
\begin{gather*}
a_b:=a-b.
\end{gather*}
From Proposition \ref{Pfaffian expression of SOPs} and the relation \eqref{2te}, It is easy to see
\begin{gather}\label{eeee}
q_{2n}^{s,t}(z)=\frac{\hat{\tau}_n^{s,t}}{\tau_n^{s,t}}
\end{gather}
and
\begin{gather}
 T_s[\mathrm{Pf}(i,j)]=\mathrm{Pf}(i+1,j+1)-\mu \mathrm{Pf}(i+1,j)-\mu \mathrm{Pf}(i,j+1)+\mu^2 \mathrm{Pf}(i,j),\nonumber\\
 T_t[\mathrm{Pf}(i,j)]=\mathrm{Pf}(i+1,j+1)-\lambda \mathrm{Pf}(i+1,j)-\lambda \mathrm{Pf}(i,j+1)+\lambda^2 \mathrm{Pf}(i,j).\label{disp}
\end{gather}
With the help of \eqref{linear:pfaff} and \eqref{zero:pfaff}, we can see
\begin{gather*}
 \mathrm{Pf}(0, \dots ,2n,\mu )
 =\mathrm{Pf}(0,\dots, 2n,\mu) - \mu \mathrm{Pf}(0,\dots ,2n-1,2n-1,\mu )\nonumber\\
\phantom{\mathrm{Pf}(0, \dots ,2n,\mu ) }{} =\mathrm{Pf}(0,\dots ,2n-1,\widetilde{2n-1},\mu ),
\end{gather*}
where $\widetilde{j}$ denotes $(j+1)-\mu \cdot j$ in the sense of \eqref{linear:pfaff}.
Iterating this transformation, we can f\/ind
\begin{gather}\label{bbbb}
\mathrm{Pf}(0, \dots ,2n,\mu )=\mathrm{Pf}(0,\widetilde{0},\widetilde{1},\dots , \widetilde{2n-1},\mu ).
\end{gather}
Meanwhile, we can also f\/ind
\begin{gather}
 \mathrm{Pf}(0,\widetilde{i_0},\dots ,\widetilde{i_{2n-1}},\mu)\nonumber\\
 \qquad{}
 = \mathrm{Pf}(\widetilde{i_0},\dots ,\widetilde{i_{2n-1}})-\sum_{k=0}^{2n-1}(-1)^k\mathrm{Pf}(\widetilde{i_k},\mu )\mathrm{Pf}(0,\widetilde{i_0},\dots ,\widehat{\widetilde{i_k}},\dots ,\widetilde{i_{2n-1}},\mu ),
\label{bb}
\end{gather}
where we have used the Pfaf\/f\/ian expansion \eqref{anotherPfaff}.
With the help of  Lemma~\ref{lemma1} and \eqref{disp}, it is straightforward to see
\begin{gather*}
 \mathrm{Pf}(\widetilde{i},\widetilde{j})=T_s[\mathrm{Pf}(i,j)] ,\qquad \mathrm{Pf}(\widetilde{i},\mu )=0.
\end{gather*}
Therefore, combining \eqref{bbbb} and \eqref{bb}, we arrive at
\begin{gather*}
T_s[\mathrm{Pf}(0,\dots ,2n-1)]=\mathrm{Pf}(0, \dots ,2n,\mu ).
\end{gather*}
In a similar manner, by using $\mathrm{Pf}(\widetilde{i},\lambda )=(\lambda  -\mu )\lambda ^i$, we further obtain
\begin{gather*}
\mathrm{Pf}(0,\dots ,2n+1, \mu ,\lambda ) =\lambda _{\mu }T_s[\mathrm{Pf}(0,\dots ,2n,\lambda )].
\end{gather*}
Finally, arranging these results, we have
\begin{gather}
\tau_n^{s+1,t} =\mathrm{Pf}(0, \dots ,2n,\mu) , \qquad
\tau _n^{s,t+1} =\mathrm{Pf}(0,\dots ,2n,\lambda ),\nonumber\\
\tau_n^{s+1,t+1} =\lambda _{\mu} ^{-1}\mathrm{Pf}(0,\dots ,2n+1,\mu,\lambda ).\label{fff}
\end{gather}
From these results, we can easily see that
$\hat{\tau }_{n}^{s,t}$ is equal to $\tau_n^{s+1,t}$ with the substitution $z=\lambda$.
Hence, in a similar manner, we can also obtain
\begin{gather}
\hat{\tau}_n^{s+1,t} =z_{\mu}  ^{-1}\mathrm{Pf}(0,\dots ,2n+1,\mu ,z),\qquad
\hat{\tau}_n^{s,t+1} =z_{\lambda } ^{-1}\mathrm{Pf}(0,\dots ,2n+1,\lambda ,z),\nonumber\\
\hat{\tau}_n^{s+1,t+1} =z_{\mu } ^{-1}z_{\lambda} ^{-1}\lambda _{\mu} ^{-1}\mathrm{Pf}(0,\dots ,2n+2,\mu ,\lambda ,z).\label{ggg}
\end{gather}
Finally, from \eqref{eeee}, \eqref{fff} and \eqref{ggg}, we obtain the relation \eqref{tesops}. This completes the proof.
\end{proof}

Using Theorem \ref{pfexpress}, we can obtain the following contiguous relations of even-degree SOPs.
\begin{theorem}\label{skewlax:theorem}
The series of even degree SOPs $\{ q_{2n}^{s,t}(z)\}_{n=0}^{\infty }$ satisfy the following contiguous relations:
\begin{gather}
 z_{\lambda } q_{2n}^{s,t+1}(z)-z_{\mu }q_{2n}^{s+1,t}(z)=z_{\mu,\lambda }A_{n}^{s,t}q_{2n-2}^{s+1,t+1}(z)-B_n^{s,t}q_{2n}^{s,t}(z), \nonumber\\
 z_{\mu ,\lambda } q_{2n}^{s+1,t+1}(z)-q_{2n+2}(z)=z_{\lambda }C_n^{s,t}q_{2n}^{s,t+1}(z)-z_{\mu }D_n^{s,t} q_{2n}^{s+1,t},\label{slax}
\end{gather}
with
\begin{alignat}{3}
& A_{n}^{s,t} =(\mu -\lambda )\frac{\tau _{n+1}^{s,t}\tau _{n-1}^{s+1,t+1}}{\tau _{n}^{s+1,t}\tau _{n}^{s,t+1}},\qquad&&  B_{n}^{s,t}=(\mu -\lambda )\frac{\tau _{n}^{s,t}\tau _{n}^{s+1,t+1}}{\tau _{n}^{s+1,t}\tau _{n}^{s,t+1}},&\nonumber\\
& C_{n}^{s,t} =(\mu -\lambda )^{-1}\frac{\tau _{n+1}^{s+1,t}\tau _{n}^{s,t+1}}{\tau _{n+1}^{s,t}\tau _{n}^{s+1,t+1}},\qquad&&  D_{n}^{s,t}=(\mu -\lambda )^{-1}\frac{\tau _{n+1}^{s,t+1}\tau _{n}^{s+1,t}}{\tau _{n+1}^{s,t}\tau _{n}^{s+1,t+1}},& \label{taupfaff}
\end{alignat}
where $\tau_n^{s,t}$ is defined in \eqref{tau}.
\end{theorem}
\begin{proof}
Using the identities of Pfaf\/f\/ians \cite{Knuth}:
\begin{gather}
 \mathrm{Pf}(i_0,\dots ,i_{2n-1},a,b,c,d)\mathrm{Pf}(i_0,\dots ,i_{2n-1})
 =\mathrm{Pf}(i_0,\dots ,i_{2n-1},a,b)\mathrm{Pf}(i_0,\dots ,i_{2n-1},c,d)\nonumber\\
\qquad{} -\mathrm{Pf}(i_0,\dots ,i_{2n-1},a,c)\mathrm{Pf}(i_0,\dots ,i_{2n-1},b,d)\nonumber\\
\qquad{} +\mathrm{Pf}(i_0,\dots ,i_{2n-1},a,d)\mathrm{Pf}(i_0,\dots ,i_{2n-1},b,c)\label{pfid1}
\end{gather}
and
\begin{gather}
 \mathrm{Pf}(i_0,\dots ,i_{2n},a,b,c)\mathrm{Pf}(i_0,\dots ,i_{2n},d)
 =\mathrm{Pf}(i_0,\dots ,i_{2n},a,b,d)\mathrm{Pf}(i_0,\dots ,i_{2n},c)\nonumber\\
\qquad{} -\mathrm{Pf}(i_0,\dots ,i_{2n},a,c,d)\mathrm{Pf}(i_0,\dots ,i_{2n},b)\nonumber\\
\qquad{} +\mathrm{Pf}(i_0,\dots ,i_{2n},b,c,d)\mathrm{Pf}(i_0,\dots ,i_{2n},a),\label{pfid2}
\end{gather}
we obtain the relation between $\tau_n^{s,t}$ and $\hat{\tau}_n^{s,t}$:
\begin{gather}
z_{\mu,\lambda}\lambda _{\mu}\tau_{n+1}^{s,t}\hat{\tau} _{n}^{s+1,t+1} =z_{\lambda }\tau_{n+1}^{s+1,t}\hat{\tau}_{n}^{s,t+1}-z_{\mu}\tau_{n+1}^{s,t+1}\hat{\tau}_n^{s+1,t}+\lambda _{\mu }\hat{\tau}_{n+1}^{s,t}\tau_n^{s+1,t+1},
\nonumber\\
\lambda _{\mu}\tau_n^{s+1,t+1}\hat{\tau}_n^{s,t} =z _{\mu }\tau_n^{s,t+1}\hat{\tau }_n^{s+1,t}-z_{\lambda }\tau_n^{s+1,t}\hat{\tau }_n^{s,t+1}+z_{\mu,\lambda }\tau_{n+1}^{s,t}\hat{\tau }_{n-1}^{s+1,t+1},\label{dckp}
\end{gather}
where $z_{\mu,\lambda }=z_{\mu}z_{\lambda }$.
Hence, from Theorem~\ref{pfexpress}, we can easily arrive at the result.
\end{proof}

Observing Theorem~\ref{skewlax:theorem}, we can regard the relation \eqref{slax} as the discrete Lax pair whose wave function is the sequence of even-degree SOPs.
Then we can directly obtain the following system.
\begin{theorem}
From the compatibility condition of \eqref{slax}, we can derive the following nonlinear difference system:
\begin{gather}
 A_n^{s+1,t+1}-A_{n+1}^{s,t}+B_{n+1}^{s,t}-B_n^{s+1,t+1}  =C_n^{s,t+1}-C_n^{s+1,t}+D_n^{s+1,t}-D_n^{s,t+1},\nonumber\\
 A_n^{s+1,t}C_{n-1}^{s+1,t}=A_n^{s,t}C_n^{s,t},\qquad A_n^{s,t+1}D_{n-1}^{s,t+1}=A_n^{s,t}D_n^{s,t},\nonumber\\
 B_n^{s+1,t}D_n^{s+1,t}=B_{n+1}^{s,t}D_n^{s,t},\qquad B_n^{s,t+1}C_n^{s,t+1}=B_{n+1}^{s,t}C_n^{s,t}.\label{dpfl}
\end{gather}
This system has a special solution which can be written in terms of Pfaffians as \eqref{taupfaff}.
\end{theorem}
We can consider the system \eqref{dpfl} as a discrete integrable system since this admits a discrete Lax pair by construction and we can put plenty arbitrary parameters into the solution \eqref{taupfaff}.

From the point of view of the $\tau $-function $\tau_n^{s,t}$, the relation \eqref{disp} which the elements of the Pfaf\/f\/ian satisfy is the same as the dispersion relation of the discrete coupled KP (dcKP) equation~\cite{Gilson} and indeed one can notice that the equation~\eqref{dckp} is just the same as the dcKP equation. Under the choice of $z$ as the spectral parameter, the dcKP equation yields the discrete Lax pair~\eqref{slax}. Therefore the system~\eqref{dpfl} can be regarded as the nonlinear form of the dcKP.

\section[An extension to $2\times 2$ matrix form]{An extension to $\boldsymbol{2\times 2}$ matrix form}\label{section4}

In the previous section, we propose the discrete integrable system in $(2+1)$-dimensional form~\eqref{dpfl} from the discrete spectral transformation of the SOPs.
This system, by construction, is related only to the even-degree SOPs and not to odd-degree SOPs.
Taking it into account that the integrable systems associated with both even and odd degree SOPs were already derived~\cite{Adler2,Leur}, we can expect that
there should exist a discrete integrable system in $2+1$ dimension which is also connected to the discrete spectral transformations of both even and odd degree SOPs.
In order to construct such system, as is described in the previous section, the ambiguity of odd-degree SOPs \eqref{ambiguity} is the barrier.
Conversely, in this section, we shall construct such system not avoiding the ambiguity of odd-degree SOPs but utilizing this.

Using the skew-inner product $\langle \cdot |\cdot \rangle^{s,t}$ with the property \eqref{2te}, we introduce the new sequence of polynomials $\{ \phi _n^{s,t}(z) \}_{n=0}^{\infty }$:
\begin{gather}\label{oesop}
\phi_{2n}^{s,t}(z)=\frac{\hat{\tau _n}^{s,t}}{\sigma_n^{s,t}},\qquad \phi_{2n+1}^{s,t}(z)=\frac{\hat{\sigma}_n^{s,t}}{\tau_n^{s,t}},
\end{gather}
where $\sigma _n^{s,t}$ and $\hat{\sigma }_n^{s,t}$ are def\/ined by
\begin{gather}
\sigma_n^{s,t}:=\mathrm{Pf}(0,\dots ,2n-2,2n)+(s\mu +t\lambda )\tau_n^{s,t},\nonumber\\
\hat{\sigma} _n^{s,t}:=\mathrm{Pf}(0,\dots ,2n-1,2n+1,z)+(s\mu +t\lambda )\hat{\tau}_n^{s,t},\label{ntau}
\end{gather}
respectively. Recall that $\tau_n^{s,t}$ and $\hat{\tau}_{n}^{s,t}$ are def\/ined by~\eqref{tau}.
here we remark that $\phi _{2n}^{s,t}(z)$ are no longer monic except for $s=t=0$.
Nevertheless, from Proposition \ref{Pfaffian expression of SOPs}, one can easily f\/ind $\{ \phi _n^{s,t}(z)\}_{n=0}$ are also SOPs with respect to $\langle \cdot | \cdot \rangle^{s,t}$.

As is seen in the previous section, in order to derive the discrete Lax pair of even-degree SOPs, that is the relation between $\tau_n^{s,t}$ and $\hat{\tau}_n^{s,t}$, the equation \eqref{dckp} plays a central role.
According to this observation, we wish to acquire the relation between $\tau_n^{s,t}$, $\hat{\tau}_n^{s,t}$, $\sigma_n^{s,t}$ and $\hat{\sigma}_n^{s,t}$.  First we examine the properties of $\sigma_n^{s,t}$ and $\hat{\sigma}_n^{s,t}$.
\begin{lemma}\label{sigmaa}
The function $\sigma_n^{s,t}$ and $\hat{\sigma}_n^{s,t}$ defined in \eqref{ntau} hold the following properties:
\begin{gather*}
\sigma_n^{s+1,t} =\mathrm{Pf}(0,\dots ,2n-1,2n+1,\mu )+(s\mu +t\lambda )\tau_n^{s+1,t},\nonumber\\
\sigma_n^{s,t+1} =\mathrm{Pf}(0,\dots ,2n-1,2n+1,\lambda )+(s\mu +t\lambda )\tau_n^{s,t+1},\nonumber\\
\sigma_n^{s+1,t+1} =\lambda _{\mu }^{-1}\mathrm{Pf}(0,\dots ,2n,2n+2,\mu ,\lambda )+(s\mu +t\lambda )\tau_n^{s+1,t+1},\nonumber\\
\hat{\sigma}_n^{s+1,t} =z_{\mu}^{-1}\mathrm{Pf}(0,\dots ,2n,2n+2,\mu,z) +(s\mu +t\lambda )\hat{\tau}_n^{s+1,t},\nonumber\\
\hat{\sigma}_n^{s,t+1} =z_{\lambda}^{-1}\mathrm{Pf}(0,\dots ,2n,2n+2,\lambda,z) +(s\mu +t\lambda )\hat{\tau}_n^{s,t+1},\nonumber\\
\hat{\sigma}_n^{s+1,t+1} =\lambda_{\mu}^{-1}z_{\mu,\lambda }^{-1}\mathrm{Pf}(0,\dots ,2n+1,2n+3,\mu,\lambda, z) +(s\mu +t\lambda )\hat{\tau}_n^{s+1,t+1}.
\end{gather*}
\end{lemma}
\begin{proof}
In a similar manner to the proof of Theorem~\ref{pfexpress}, we can see
\begin{gather*}
\mathrm{Pf}(0,\dots, \widehat{k+1},\dots ,2n+1,\mu ) =\mathrm{Pf}(0,\dots ,k,k+2,\widetilde{k+2},\dots ,\widetilde{2n},\mu )\nonumber\\
\qquad{} =\mathrm{Pf}(0,\dots ,k,k+2,\widetilde{k+2},\dots ,\widetilde{2n},\mu )-\mu^2 \mathrm{Pf}(0,\dots ,k,k,\widetilde{k+2},\dots ,\widetilde{2n},\mu )
\nonumber\\
\qquad{} =\mathrm{Pf}(0,\dots k,(k+2)-\mu^2 \cdot k,\widetilde{k+2},\dots ,\widetilde{2n},\mu )\nonumber\\
\qquad{} =\mathrm{Pf}(0,\widetilde{0},\dots \widetilde{k},(k+2)-\mu^2\cdot k,\widetilde{k+2},\dots ,\widetilde{2n},\mu ).
\end{gather*}
Noticing $(k+2)-\mu^2\cdot k=\widetilde{k+1}-\mu \cdot \widetilde{k}$, we get
\begin{gather*}
 \mathrm{Pf}(0,\dots, \widehat{k+1},\dots ,2n+1,\mu )
 =\mathrm{Pf}(0,\widetilde{0},\dots, \widehat{\widetilde{k}},\dots ,\widetilde{2n},\mu )-\mu \mathrm{Pf}(0,\widetilde{0},\dots, \widehat{\widetilde{k+1}},\dots ,\widetilde{2n},\mu )\nonumber\\
 \qquad{} =T_s[\mathrm{Pf}(0,\dots ,\widehat{k},\dots ,2n-1)]-\mu T_s[\mathrm{Pf}(0,\dots ,\widehat{k+1},\dots ,2n-1)],
\end{gather*}
where we have used the relation \eqref{bb}.
Therefore  with the symmetry $\mu \leftrightarrow \lambda $, we have
\begin{gather*}
T_s[\mathrm{Pf}(0,1,\dots ,2n-2,2n)] =\mathrm{Pf}(0,1,\dots ,2n-1,2n+1,\mu )-\mu \tau_n^{s+1,t},\nonumber\\
T_t[\mathrm{Pf}(0,1,\dots ,2n-2,2n)] =\mathrm{Pf}(0,1,\dots ,2n-1,2n+1,\mu )-\lambda \tau_n^{s,t+1}.
\end{gather*}
In a similar manner, we can also obtain
\begin{gather*}
 T_s\circ T_t[\mathrm{Pf}(0,1,\dots ,2n-2,2n)]
 =\lambda _{\mu}^{-1} \mathrm{Pf}(0,1,\dots ,2n-1,2n+1,\mu )-(\mu +\lambda )\tau_n^{s+1,t+1},\nonumber\\
 T_s[\mathrm{Pf}(0,1,\dots ,2n-1,2n+1,z)]
 =z _{\mu}^{-1} \mathrm{Pf}(0,1,\dots ,2n,2n+2,\mu, z )-\mu\hat{\tau}_n^{s+1,t},\nonumber\\
 T_t[\mathrm{Pf}(0,1,\dots ,2n-1,2n+1,z)]
 =z _{\lambda }^{-1} \mathrm{Pf}(0,1,\dots ,2n,2n+2,\lambda , z )-\lambda \hat{\tau}_n^{s,t+1}, \nonumber\\
 T_s\circ T_t[\mathrm{Pf}(0,1,\dots ,2n-1,2n+1,z)]\nonumber\\
\qquad{} =z _{\mu,\lambda }^{-1} \mathrm{Pf}(0,1,\dots ,2n+1,2n+3,\mu, \lambda ,z )-(\mu +\lambda )\hat{\tau}_n^{s+1,t+1}.
\end{gather*}
These results immediately lead us to the conclusion.
\end{proof}

By virtue of Lemma~\ref{sigmaa}, we can f\/ind the relation between  $\tau_n^{s,t}$, $\hat{\tau}_n^{s,t}$, $\sigma_n^{s,t}$ and $\hat{\sigma}_n^{s,t}$.
\begin{theorem}
The function $\tau_n^{s,t}$, $\hat{\tau}_n^{s,t}$, $\sigma_n^{s,t}$ and $\hat{\sigma}_n^{s,t}$ defined in \eqref{tau} and \eqref{ntau} satisfy the following four relations:
\begin{gather}
z_{\mu,\lambda }\lambda _{\mu } \sigma_{n+1}^{s,t}\hat{\tau}_{n-1}^{s+1,t+1} =z_{\lambda }\tau_n^{s+1,t}\hat{\sigma}_n^{s,t+1}-z_{\mu }\tau_n^{s,t+1}\hat{\sigma }_n^{s+1,t}+\lambda _{\mu }\sigma_n^{s+1,t+1}\hat{\tau}_n^{s,t},\nonumber\\
z_{\mu,\lambda }\lambda _{\mu } \tau_{n+1}^{s,t}\hat{\sigma }_{n-1}^{s+1,t+1} =z_{\lambda }\sigma_n^{s+1,t}\hat{\tau }_n^{s,t+1}-z_{\mu }\sigma_n^{s,t+1}\hat{\tau }_n^{s+1,t}+\lambda _{\mu }\tau_n^{s+1,t+1}\hat{\sigma }_n^{s,t},\nonumber\\
z_{\mu,\lambda }\lambda _{\mu } \sigma_{n+1}^{s,t}\hat{\tau }_{n}^{s+1,t+1} =z_{\lambda }\tau_{n+1}^{s+1,t}\hat{\sigma }_n^{s,t+1}-z_{\mu }\tau_{n+1}^{s,t+1}\hat{\sigma }_n^{s+1,t}+\lambda _{\mu }\sigma_n^{s+1,t+1}\hat{\tau }_{n+1}^{s,t},\nonumber\\
z_{\mu,\lambda }\lambda _{\mu } \tau_{n+1}^{s,t}\hat{\sigma }_{n}^{s+1,t+1} =z_{\lambda }\sigma_{n+1}^{s+1,t}\hat{\tau }_n^{s,t+1}-z_{\mu }\sigma_{n+1}^{s,t+1}\hat{\tau }_n^{s+1,t}+\lambda _{\mu }\tau_n^{s+1,t+1}\hat{\sigma }_{n+1}^{s,t}.\label{edckp}
\end{gather}
\end{theorem}
\begin{proof}
Pfaf\/f\/ian identities \eqref{pfid1}, \eqref{pfid2} and Lemma~\ref{sigmaa} directly lead us to the~\eqref{edckp}.
This completes the proof.
\end{proof}

If we take $\sigma =\tau $, one f\/inds that the equations~\eqref{edckp} go to the discrete coupled KP equation~\eqref{dckp}. Therefore, these equations can be considered as the natural extension of the discrete coupled KP equation from the point of view of the SOPs. Hereafter we will call~\eqref{dckp} as the extended discrete coupled KP (edcKP) equation.

The edcKP equation gives us the contiguous relation among $\{ \phi _n^{s,t}(z)\}$:
\begin{theorem}
Let $\Phi_n^{s,t}(z)$ be the vector function given by
\begin{gather*}
\Phi_n^{s,t}(z)=
\begin{pmatrix}
\phi_{2n}^{s,t}(z)\\
\phi_{2n+1}^{s,t}(z)
\end{pmatrix}.
\end{gather*}
The vector functions $\{ \Phi_n^{s,t}(z)\}_{n=0}^{\infty }$ satisfy the following two vector linear equations:
\begin{gather}
 z_{\lambda } \Phi_n^{s,t+1}(z)-z_{\mu }\Phi_n^{s+1,t}(z)=z_{\mu,\lambda }\mathbf{A}_{n}^{s,t}\Phi_{n-1}^{s+1,t+1}(z)-\mathbf{B}_n^{s,t}\Phi_{n}^{s,t}(z), \nonumber\\
 z_{\mu ,\lambda } \Phi_{n}^{s+1,t+1}(z)-\Phi_{n+1}(z)=z_{\lambda }\mathbf{C}_n^{s,t}\Phi_{n}^{s,t+1}(z)-z_{\mu }\mathbf{D}_n^{s,t} \Phi_{n}^{s+1,t},\label{edlax}
\end{gather}
where the coefficients $\mathbf{A}_n^{s,t}$, $\mathbf{B}_n^{s,t}$, $\mathbf{C}_n^{s,t}$, $\mathbf{D}_n^{s,t}$ can be written explicitly in terms of Pfaffians:
\begin{gather}
 \mathbf{A}_n^{s,t}=(\lambda -\mu )
\begin{pmatrix}
0 & \dfrac{\tau_{n+1}^{s,t}\tau_{n-1}^{s+1,t+1}}{\sigma_n^{s+1,t}\sigma_n^{s,t+1}}\vspace{1mm}\\
\dfrac{\sigma_{n+1}^{s,t}\sigma_{n-1}^{s+1,t+1}}{\tau_n^{s+1,t}\tau_n^{s,t+1}} & 0
\end{pmatrix}, \nonumber\\
 \mathbf{B}_n^{s,t}=(\lambda -\mu )
\begin{pmatrix}
0 & \dfrac{\tau_{n}^{s,t}\tau_{n}^{s+1,t+1}}{\sigma_n^{s+1,t}\sigma_n^{s,t+1}}\vspace{1mm}\\
\dfrac{\sigma_{n}^{s,t}\sigma_{n}^{s+1,t+1}}{\tau_n^{s+1,t}\tau_n^{s,t+1}} & 0
\end{pmatrix},\nonumber\\
 \mathbf{C}_n^{s,t}=(\lambda -\mu )^{-1}
\begin{pmatrix}
0 & \dfrac{\tau_{n+1}^{s+1,t}\tau_{n}^{s,t+1}}{\sigma_{n+1}^{s,t}\sigma_n^{s+1,t+1}}\vspace{1mm}\\
\dfrac{\sigma_{n+1}^{s+1,t}\sigma_{n}^{s,t+1}}{\tau_{n+1}^{s,t}\tau_n^{s+1,t+1}} & 0
\end{pmatrix},\nonumber\\
 \mathbf{D}_n^{s,t}=(\lambda -\mu )^{-1}
\begin{pmatrix}
0 & \dfrac{\tau_{n+1}^{s,t+1}\tau_{n}^{s+1,t}}{\sigma_{n+1}^{s,t}\sigma_n^{s+1,t+1}}\vspace{1mm}\\
\dfrac{\sigma_{n+1}^{s,t+1}\sigma_{n}^{s+1,t}}{\tau_{n+1}^{s,t}\tau_n^{s+1,t+1}} & 0
\end{pmatrix}.\label{esol}
\end{gather}
\end{theorem}
\begin{proof}
The def\/initions of $\{ \phi _n^{s,t}\}_{n=0}^{\infty } $ \eqref{oesop} and the edcKP equation \eqref{edckp} directly lead us to the conclusion. This completes the proof.
\end{proof}

We can regard \eqref{edlax} as the discrete vector Lax pair of SOPs of both even and odd degree. Therefore the discrete Lax pair shows us the following systems associated with SOPs of both even and odd degree,  which can be regarded as a discrete integrable system along with \eqref{dpfl}:
\begin{theorem}
From the compatibility condition of the discrete vector Lax pair \eqref{edlax}, we can derive the nonlinear system in $2+1$ dimension which takes the $2\times 2$ matrix form:
\begin{gather}
 \mathbf{A}_{n}^{s+1,t+1}-\mathbf{A}_{n+1}^{s,t}+\mathbf{B}_{n+1}^{s,t}-\mathbf{B}_n^{s+1,t+1}
  =\mathbf{C}_n^{s,t+1}-\mathbf{C}_n^{s+1,t}+\mathbf{D}_n^{s+1,t}-\mathbf{D}_n^{s,t+1},\nonumber\\
 \mathbf{A}_n^{s+1,t}\mathbf{C}_{n-1}^{s+1,t}=\mathbf{A}_n^{s,t}\mathbf{C}_{n}^{s+1,t},\qquad
\mathbf{A}_n^{s,t+1}\mathbf{D}_{n-1}^{s,t+1}=\mathbf{A}_n^{s,t}\mathbf{D}_n^{s,t},\nonumber\\
 \mathbf{B}_n^{s+1,t}\mathbf{D}_n^{s+1,t}=\mathbf{B}_{n+1}^{s,t}\mathbf{D}_n^{s,t},\qquad
\mathbf{B}_n^{s,t+1}\mathbf{D}_n^{s,t+1}=\mathbf{B}_{n+1}^{s,t}\mathbf{D}_n^{s,t}.\label{edpfl}
\end{gather}
Special solutions of the nonlinear system~\eqref{edpfl} can be written in terms of Pfaffians given by~\eqref{esol}.
\end{theorem}

One can easily f\/ind that this system is the $2\times 2$ matrix extension of the system~\eqref{edpfl} and this is the nonlinear form of the edcKP equation. This system is clearly connected to the SOPs of both even and odd degree and from its construction, we can regard the system~\eqref{edpfl} as the discrete coutnerpart of the Pfaf\/f lattice in $(2+1)$-dimensional form.

\section{Concluding remarks}\label{section5}

In this paper, we proposed the discrete spectral transformation of SOPs.
From the discrete spectral transformation of SOPs, we derived the discrete integrable systems both in $(1+1)$-dimensional form and in $(2+1)$-dimensional form. In $(1+1)$-dimensional case, the system gives the integrable discretization of the Pfaf\/f lattice, though the system is of course not local.  In $(2+1)$-dimensional case, the system is local and corresponds only to the even-degree SOPs. This system is also shown to be connected to discrete coupled KP equation. The discrete coupled KP equation is known to be equivalent to the discretized Hirota--Ohta equation \cite{Hirota} in the bilinear form levels. Recently, the discrete integrable system is proposed from the auxiliary linear problems for the Hirota--Ohta (or DKP) system \cite{Adler4,Takasaki}.

We further extended the discrete integrable system associated with even-degree SOPs to the system associated with both even and odd degree SOPs. This system can be described in $2\times 2$ matrix form, which is related to an extension of the discrete coupled KP equation. This extension has never been considered before, although it is quite natural from the aspects of SOPs.

Finally, let us mention for the relationship between our results and the random matrix theory.  As a by-product of the discrete transformation of SOPs, we found the compact factorization of the kernel function of the correlation function for the orthogonal or symplectic ensembles in terms of the two even-degree SOPs with respect to the dif\/ferent weights.
This factorization is not known as far as the authors know and indeed might be a new one.
Moreover, it is known that a matrix integral over orthogonal and symplectic  ensembles of Hermitian matrices can be considered as a~$\tau$-function of the coupled KP hierarchy~\cite{Kakei}, while
our discrete integrable systems~\eqref{dpfl} and~\eqref{edpfl} are directly connected to discrete coupled KP and its extension, respectively.  We hope these results will bring a deeper understanding of the random matrix theory and several applications in its works in the near future.

\subsection*{Acknowledgement}

The authors would like to thank the Centre de recherches math\'{e}matiques (CRM) for its hospitality.
The work of H.M.\  is partially supported by a Grant-in-Aid for Japan Society for the Promotion of Science (JSPS) Fellows.
The research of S.T.\ is supported in part by KAKENHI (22540224).
The authors should like to thank the referees for the careful reading of the manuscript and a lot of helpful suggestions and comments.

\pdfbookmark[1]{References}{ref}
\LastPageEnding

\end{document}